\documentclass[a4paper,10pt]{article}
\usepackage{amsmath,amssymb,epsfig,graphicx} 

\begin{document}

\title{A consistent event generation for non-resonant diphoton production at hadron collisions}

\author{Shigeru Odaka\\
High Energy Accelerator Research Organization (KEK)\\
1-1 Oho, Tsukuba, Ibaraki 305-0801, Japan\\
E-mail: \texttt{shigeru.odaka@kek.jp}}

\maketitle

\begin{abstract}
We have developed a Monte Carlo event generator for non-resonant diphoton ($\gamma\gamma$) 
production at hadron collisions in the framework of GR@PPA, 
which consistently includes processes having additional one jet radiation.
The possible double count problem in the generation of radiative processes is avoided 
by using the LLL subtraction method that we have applied to the weak-boson production processes.
The subtraction method has been extended to the final-state QED divergence that appears 
in the $qg \rightarrow \gamma\gamma + q$ process.
Because a parton shower (PS) which regularizes the subtracted QED divergence is still under development, 
we tried to use PYTHIA for the generation of the fragmentation events 
to restore the subtracted components.
The simulation employing the "old" PS of PYTHIA shows a reasonable matching with the GR@PPA events, 
and the combined event sample shows a result in reasonable agreement with ResBos.
We found that the contribution from $qg \rightarrow \gamma\gamma + q$ is significant in the LHC condition.
This event generator must be useful for the background studies in low-mass Higgs boson searches at LHC.
\end{abstract}

\section{Introduction}

Diphoton ($\gamma\gamma$) production is one of most promising channels for the discovery 
of the Higgs boson having a relatively small invariant mass ($\lesssim 130$ GeV/$c^{2}$) at LHC. 
However, the measurement must suffer from a large irreducible diphoton background produced 
via non-resonant QED interactions.
It is necessary to understand the properties of this background, 
not only for the discovery of the Higgs boson but also for detailed studies after the discovery.
The identification of photons is rather complicated and largely dependent on the detector performance.
For instance, a certain isolation condition has to be required in order to reduce large contamination 
of $\pi^{0}$ from hadron jets. 
The performance of this selection depends on the details of detector responses 
and is hard to be evaluated analytically.
Therefore, it is strongly desired to provide theoretical predictions in the form of Monte Carlo (MC) event generators.

\begin{figure}[t]
\centerline{\psfig{file=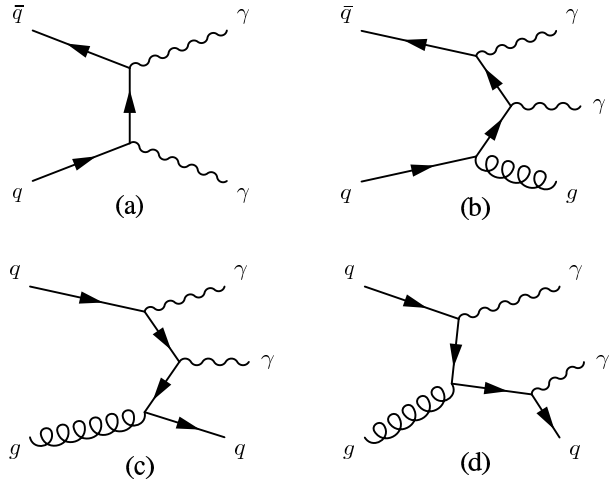,width=100mm}}
\caption{Typical Feynman diagrams for non-resonant diphoton production at hadron collisions: 
(a) the lowest order, (b) a gluon radiation process, 
(c) a quark radiation process, and (d) another quark radiation process.
The processes (b) and (c) have initial-state QCD divergences, while (d) has a final-state QED divergence.
\protect\label{fig:diagrams}}
\end{figure}

The lowest order process for non-resonant diphoton production is very simple as shown in Fig.\ \ref{fig:diagrams} (a). 
Despite that, the next-to-leading order (NLO) correction to this process is known to be very large \cite{Binoth:1999qq,Balazs:2007hr}.
The large correction is predominantly due to the contribution from real radiation processes 
illustrated in Fig.\ \ref{fig:diagrams} (b) to (d).
While the contribution from gluon radiation processes shown in Fig.\ \ref{fig:diagrams} (b) is not very significant, 
those from quark radiation processes shown in Fig.\ \ref{fig:diagrams} (c) and (d) may become large 
due to a very large gluon density inside protons.
It must be necessary to include these processes in order to provide a realistic simulation.

In this report, we describe the MC event generator for non-resonant diphoton production at hadron collisions that we have developed.
The program has been developed in the framework of the GR@PPA event generator \cite{Tsuno:2006cu}, 
and supports the generation of radiative processes in Fig.\ \ref{fig:diagrams} (b) to (d) together 
with the lowest-order process in Fig.\ \ref{fig:diagrams} (a).
Though the generator includes radiative processes, it is not fully including NLO corrections. 
Non-divergent terms in soft/virtual corrections are yet to be included.
In any case, the radiative processes have various divergences which we need to regularize.
The initial-state QCD divergences can be regularized 
using the method that we have applied to weak-boson production processes, 
where divergent terms are numerically subtracted from the matrix elements of radiative processes 
(the LLL subtraction) \cite{Kurihara:2002ne,Odaka:2007gu,Odaka:2009qf}.
The subtracted contributions are restored by combining with non-radiative processes 
to which a parton shower (PS) is applied.
We can avoid the double count problem by the subtraction and naturally regularize the divergences 
as a result of the multiple radiation in PS.

The quark radiation process illustrated in Fig.\ \ref{fig:diagrams} (c) and (d) not only has an initial-state QCD divergence 
but also has a QED divergence in the final state.
We have extended the method applied to the initial-state QCD to this final-state QED divergence.
The extension of the subtraction is straightforward, 
while the preparation of an appropriate PS is not trivial since it has to support QED radiations together with QCD.
This PS has to be applied to the quark in the final state of the $qg \rightarrow \gamma + q$ process 
to radiate photons from the final-state quark based on a collinear approximation (the fragmentation process).
It is desired to have a capability to force a hard photon radiation in this PS 
since we are interested in those events having two hard photons in the final state.
Such a final-state PS is still in development. 
Instead, we try to use the PYTHIA PS \cite{Sjostrand:2006za} for simulating the fragmentation process in this report.
Though the PYTHIA PS is capable of radiating photons, it does not have a mechanism to force a hard photon radiation.
We need to repeat the generation until we observe a sufficiently hard photon in the generated event.
We use the so-called "old" PS in the present study.
The "new" PS is not used because we still have some questions on its behavior.

We sometimes find reports in which the detection efficiency and acceptance 
for the diphoton measurements are evaluated 
using event generators for the lowest-order process and fragmentation processes.
Such evaluations are not self-consistent.
Parton showers used in the simulation of fragmentation processes have a certain energy scale 
defining the maximum hardness of the parton radiation.
The results depend on this arbitrarily chosen energy scale since those radiations exceeding this scale are ignored.
Non-collinear contributions are also ignored.
These ignored contributions may be small in many processes compared to the contributions taken into consideration.
However, as we will show in this report, they become comparable to the lowest-order contribution 
in the diphoton production. 
It is necessary to include a simulation based on the exact matrix elements for the $qg \rightarrow \gamma\gamma + q$ process 
in order to make a reliable evaluation.

The fragmentation process that we take into consideration in this report is the so-called "single" fragmentation.
The "double" fragmentation in which two photons are radiated from final-state partons, for instance, 
from quarks in the $gg \rightarrow q\bar{q}$ process are not supported.
We need to introduce "$\gamma\gamma$ + 2 jets" production processes 
in order to construct a consistent event generator including the "double" fragmentation.
Besides, $gg \rightarrow \gamma\gamma$ and its higher orders are not included at present.

We require a typical kinematical condition for the Higgs-boson search at LHC through the present study 
because we are interested in its background; that is
\begin{eqnarray}
  & p_{T}(\gamma_{1}) \geq 40 \text{ GeV/}c, \ 
  p_{T}(\gamma_{2}) \geq 25 \text{ GeV/}c, \nonumber \\
  & |\eta(\gamma)| \leq 2.5, \ \Delta R(\gamma\gamma) \geq 0.4 \nonumber \\
  & 80 \leq m_{\gamma\gamma} \leq 140 \text{ GeV/}c^{2}.
\label{eq:selection}
\end{eqnarray}
We apply an asymmetric requirement to the transverse momenta ($p_{T}$) of photons with respect to the incident beam direction.
The requirement on the pseudorapidity ($\eta$) is common to the two photons.
In addition, though this is not effective for real diphoton events now we consider, 
we require a sufficient $\Delta R$ separation between the two photons, 
where $\Delta R$ is defined from the differences in the pseudorapidity ($\eta$) and the azimuthal angle ($\phi$) as 
$\Delta R^{2} = \Delta\eta^{2} + \Delta\phi^{2}$.
Finally, the invariant mass of the two photons ($m_{\gamma\gamma}$) is restricted within the range that we are interested in.
These conditions are required after completing the simulation down to the hadron level.
Looser constraints are applied at the event generation in order to avoid any bias.
The simulations are carried out for the design condition of LHC, 
proton-proton collisions at a center-of-mass (cm) energy of 14 TeV.

This report is organized as follows: 
the extension of the limited leading-log (LLL) subtraction method to the final-state QED divergence is 
described in Section 2, 
and the simulation of the fragmentation process employing the PYTHIA PS is described in Section 3.
A typical isolation cut is simulated in Section 4.
The results from the combined event simulation is presented in Section 5, 
and the discussions are concluded in Section 6.

\section{Final-state QED LLL subtraction}

We approximate the final-state QED divergence in the $qg \rightarrow \gamma\gamma + q$ process as
%\begin{eqnarray}
\begin{equation}
  \left|\mathcal{M}_{qg \rightarrow \gamma\gamma q}^{(LLL,\text{fin})}
  ({\hat s}, {\hat \Phi}_{\gamma\gamma q})\right|^{2}
  = \left|\mathcal{M}_{qg \rightarrow \gamma q}({\hat s}, {\hat \Phi}_{\gamma q})\right|^{2}
  f^{(LL)}_{q \rightarrow q \gamma}(Q^{2},z) \theta(\mu_{FSR}^{2} - Q^{2}) .
\label{eq:lll}
%\end{eqnarray}
\end{equation}
The leading-log (LL)  radiation function can be given as 
\begin{equation}\label{eq:rad}
  f^{(LL)}_{q \rightarrow q \gamma}(Q^{2},z)={\alpha \over 2\pi}  
  {16 \pi^{2} \over Q^{2}} P_{q \rightarrow q \gamma}(z).
\end{equation}
The parameter $\alpha$ is the electromagnetic coupling and the splitting function is defined as 
\begin{equation}\label{eq:split}
  P_{q \rightarrow q \gamma}(z) = e_{q}^{2} {1+z^{2} \over 1-z},
\end{equation}
where $e_{q}$ is equal to 1/3 for down-type quarks and 2/3 for up-type quarks.
We evaluate Eq.\ (\ref{eq:lll}) for two possible combinations of the quark and a photon in the final state, 
and numerically subtract them from the exact matrix element of $qg \rightarrow \gamma\gamma + q$, 
together with the LLL term for the initial-state QCD divergence.
The parameters $Q^{2}$ and $z$ are defined by using the sum of the energy ($E_{q\gamma}$) 
and the momentum ($p_{q\gamma}$) of the considered $q$-$\gamma$ pair.
They are defined in the cm frame of the $qg \rightarrow \gamma\gamma + q$ event as 
\begin{equation}\label{eq:qsq}
  Q^{2} = E_{q\gamma}^{2} - p_{q\gamma}^{2},
\end{equation}
and
\begin{equation}\label{eq:z}
  z = { p_{L}( E_{q\gamma} + p_{q\gamma}) + Q^{2}/2 \over p_{q\gamma}( E_{q\gamma} + p_{q\gamma}) + Q^{2} },
\end{equation}
where $p_{L}$ is the momentum component of the $q$ in parallel to the summed momentum.
Equation\ (\ref{eq:z}) is defined so that $z$ should represent the momentum fraction 
at the infinite-momentum limit, $Q^{2}/p_{q\gamma}^{2} \rightarrow 0$.

Equation\ (\ref{eq:rad}) is slightly different from the radiation function for the initial state \cite{Kurihara:2002ne}; 
$z$ has vanished in the denominator.
This is because we assume that the cm energy does not change before and after the radiation.
The mapping to the non-radiative process, $qg \rightarrow \gamma + q$, 
which is necessary to perform in order to evaluate the LLL term in Eq.\ (\ref{eq:lll}), 
is also defined according to this assumption.
We define the momentum of the quark in the final state of the mapped $qg \rightarrow \gamma + q$ event 
by the sum of the momenta of the $q$-$\gamma$ pair in consideration.
Since all particles in the $qg \rightarrow \gamma + q$ event are assumed to be on-shell, 
the invariant mass of the final state evaluated from thus defined momentum becomes smaller than that of the initial state.
In order to compensate for this decrease, 
we increase the overall scale of the momenta of the final-state particles.
The $Q^{2}$ value defined in Eq.\ (\ref{eq:qsq}) is also increased with the same scaling factor.
The $z$ value is independent of this rescaling.
These details, the definitions in Eqs.\ (\ref{eq:qsq}) and (\ref{eq:z}) and the subsequent momentum adjustment, 
are not universal.
They have been chosen in order to achieve a good matching with the parton shower (PS) that we are developing.
Therefore, the application of the PYTHIA PS may result in a certain mismatch, 
though the effect of such details must vanish at the limit where the radiative cross section diverges.

An event generator implementing the above subtraction has been developed in the framework of GR@PPA, 
and the generation was tried for proton-proton collisions with a cm energy of 14 TeV
with CTEQ6L1 \cite{Pumplin:2002vw} used for PDF.
The energy scale was defined as
\begin{equation}\label{eq:scale}
  \mu^{2} = | \vec{p}_{T}(\gamma_{1}) - \vec{p}_{T}(\gamma_{2}) |^{2}/4,
\end{equation}
where $\vec{p}_{T}(\gamma_{i})$ denotes the transverse momentum vector of the two photons.
This definition is equivalent to the ordinary definition, $\mu = p_{T}(\gamma)$, 
for $q\bar{q} \rightarrow \gamma\gamma$.
We used the same definition for the renormalization and factorization scales.
The energy scale for the initial-state PS must be equal to the factorization scale in our method.
Though it is not necessary, we adopted the same definition for the final-state PS.
Thus, all the energy scales used in the event generation were identical in this test.

The LLL subtraction is limited by the $\theta$ function in Eq.\ (\ref{eq:lll}).
This is because the implementation of the PS to be applied to the non-radiative process 
is limited by a certain energy scale. 
Thus, the energy scale used in Eq.\ (\ref{eq:lll}), $\mu_{FSR}$, must be equal to the one 
defined for the mapped $qg \rightarrow \gamma + q$ event in order to achieve a good matching.
In the present study,  we define $\mu_{FSR}$ to be equal to the $p_{T}$ of the mapped $qg \rightarrow \gamma + q$ event.

\begin{figure}[t]
\centerline{\psfig{file=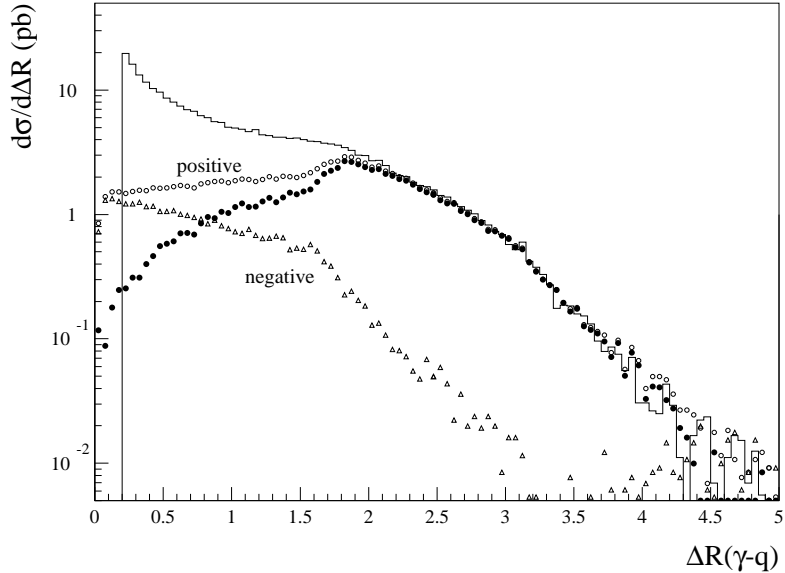,width=120mm}}
\caption{$\Delta R(\gamma\text{-}q)$ distribution of the simulated $qg \rightarrow \gamma \gamma + q$ events 
satisfying the kinematical condition in Eq.\ (\ref{eq:selection}).
The solid histogram shows the distribution before applying the final-state QED LLL subtraction, 
where $\Delta R(\gamma\text{-}q) > 0.2$ is required to cutoff the divergence.
Open circles show the distribution of positive-weight events after the subtraction, 
and open triangles show that of negative-weight events.
The final result after applying the final-state QED LLL subtraction is shown with filled circles.
The initial-state QCD LLL subtraction is already applied while parton showers are yet to be applied in this simulation.
\protect\label{fig:drgamq}}
\end{figure}

One of the simulation results is shown in Fig.\ \ref{fig:drgamq}.
In this figure, we have plotted the distribution of $\Delta R$ between the photon and the quark in the final state 
of the $qg \rightarrow \gamma \gamma + q$ events, 
where $q$ represents any quark or anti-quark up to the $b (\bar{b})$ quark.
We obtain two values since there are two photons in the final state.
We take the smaller one as the $\Delta R(\gamma\text{-}q)$.
Parton showers are yet to be applied but the initial-state QCD LLL subtraction is already applied in this simulation.
The selection condition in Eq.\ (\ref{eq:selection}) is applied to the generated events.
The distribution directly derived from the matrix element before applying the final-state LLL subtraction 
is shown with a solid histogram in the figure for comparison.
We applied a cut of $\Delta R(\gamma\text{-}q) > 0.2$ 
since this distribution is divergent at $\Delta R(\gamma\text{-}q) = 0$.

Since the LLL subtraction is unphysical, we obtain negative-weight events as well as ordinary positive-weight events 
when we apply the subtraction.
The event weights are always equal to $+1$ or $-1$ 
because BASES/SPRING \cite{Kawabata:1985yt,Kawabata:1995th} automatically unweights the events.
Therefore, we can obtain the desired distribution by subtracting the number of negative-weight events 
from that of positive-weight events in each histogram bin.
The open circles (triangles) in the figure show the distribution of positive(negative)-weight events, 
and the final distribution is shown with filled circles. 
We can see that the subtraction is effective only in a relatively small $\Delta R(\gamma\text{-}q)$ region as expected.
The distribution at large $\Delta R(\gamma\text{-}q)$ is not altered by the subtraction.
We can also see that the distributions after the subtraction converge to finite values as $\Delta R(\gamma\text{-}q) \rightarrow 0$;
not only the final result converges to zero, but also the positive and negative weight events converge to a finite value 
at this limit.
These facts show that the subtraction is done properly at least near the divergent limit.
We have applied a small cutoff, $\Delta R(\gamma\text{-}q) > 0.01$, in the event generation for numerical stability.
We can see that the effect of this cutoff is negligible.

\section{Fragmentation process by PYTHIA PS}

The subtracted LLL terms must be restored by non-radiative processes 
to which an appropriate parton shower (PS) is applied.
Because our PS is still under development, we try to use the PYTHIA PS \cite{Sjostrand:2006za} in the present study.
We used the so-called "old" PS because this model looks similar to the PS that we are developing.
We used PYTHIA 6.423 for this study.
We generated $q\bar{q} \rightarrow g \gamma$ and $qg \rightarrow q \gamma$ events by setting as 
{\tt msel = 0}, {\tt msub(14) = 1}, and {\tt msub(29) = 1} 
in the LHC condition at the design energy, 14 TeV. 
The $q\bar{q} \rightarrow g \gamma$ interaction was turned on for completeness.
Though it is very inefficient to find hard photons in the gluon fragmentation, 
turning on this process is harmless since the production cross section is small compared to $qg \rightarrow q \gamma$.

For PDF, CTEQ6L1 was applied with the help of LHAPDF/LHAGLUE \cite{Whalley:2005nh} 
in the LHAPDF 5.8.4 distribution by setting as {\tt mstp(51) = 10042} and {\tt mstp(52) = 2} in PYTHIA.
The final-state particles were allowed to produce in the region $|\eta| \leq 3.0$ 
with the transverse momenta of $p_{T} \geq 30$ GeV/$c$ at the on-shell parton level
by appropriately setting the {\tt ckin} parameters.
Though this $p_{T}$ cut may look tight compared to the condition in Eq.\ (\ref{eq:selection}), 
it is safe enough because the lower-$p_{T}$ photons are predominantly produced in the fragmentation.
Since we do not want to change the preset energy scales, 
we set the scale parameters as {\tt parp(67) = 1.0} and {\tt parp(71) = 1.0}.
All the energy scales must have been set equal to the $p_{T}$ of the on-shell parton level interaction 
with this setting.
In addition, for safety, we explicitly disabled the matrix-element correction by setting as {\tt mstp(68) = 0}.
The other parameters were left unchanged so that the default "old" PS and further simulations 
down to the hadron level should be applied according to the default setting.

The event generation with the above setting was repeated and hard photons were looked for 
in the fragmentation of the final-state partons.
The probability to find hard photons in PS is very small.
The efficiency to find events satisfying the condition in Eq.\ (\ref{eq:selection}) was $2.5 \times 10^{-4}$.
In order to improve the efficiency, we enhanced the QED radiation by a factor of 10 
by setting as {\tt mstj(41) = 10} and {\tt parj(84) = 10.0}.
The obtained results were corrected for this enhancement.
It is not recommended in the manual to apply larger enhancement factors 
because the multiple photon radiation effect may become significant.
By the way, it should be noted that the efficiency is still at the level of $10^{-3}$ even with this enhancement.

In order to investigate the matching with the subtracted GR@PPA simulation results, 
it is better to reconstruct the corresponding $qg \rightarrow \gamma\gamma + q$ events 
at the on-shell parton level from full PYTHIA simulation events in which a hard fragmentation photon is observed.
This reconstruction can be done unambiguously by using the information of the initial-state partons 
initiated the hard interaction, without using the information of the final-state partons or hadrons.
The reconstruction using the final-state partons or hadrons is dangerous because their origin 
is ambiguous.
The initial-state partons initiated the hard interaction can be found in the documentation lines 
of the PYTHIA event record as partons having no child.
We first boost the prompt and fragmentation photons to the cm frame of these partons 
in which the initial-state parton momenta are aligned along the beam direction.
We determine the momentum of the remnant on-shell parton from the photon momenta 
to balance the total momentum.
The flavor of the remnant parton can be determined from the flavor of the initial-state partons.
The total energy of the final state derived from thus determined momenta is usually different 
from the initial-state cm energy as a result of the application of PS and hadronization simulations.
We adjust the overall scale of the momenta of the final-state particles to match the total energy 
as is done in the mapping to non-radiative events in the LLL subtraction.
We then boost the reconstructed parton and photon momenta to the laboratory frame by using 
the momentum fraction information in {\tt pari(33)} and {\tt pari(34)} 
originally used for the generation of the hard interaction.

\begin{figure}[t]
\centerline{\psfig{file=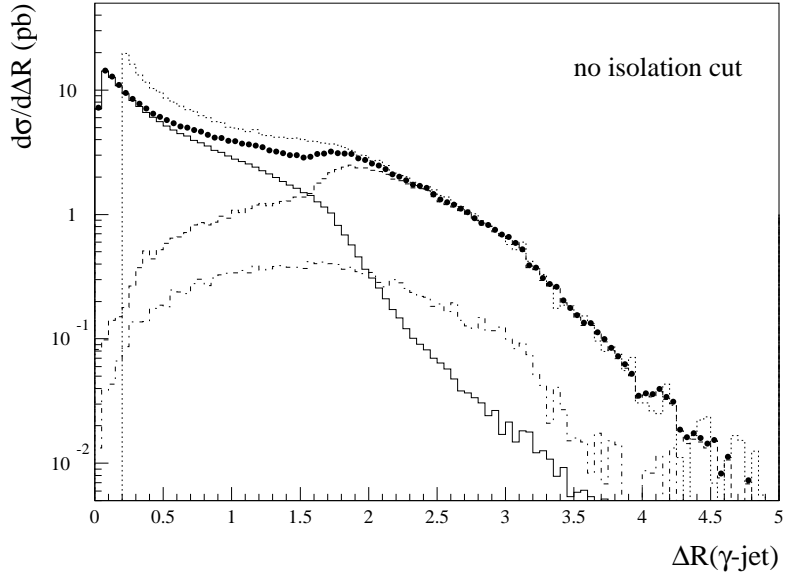,width=120mm}}
\caption{$\Delta R(\gamma\text{-jet})$ distribution of the simulated $\gamma \gamma + \text{jet}$ events 
satisfying the kinematical condition in Eq.\ (\ref{eq:selection}), 
where "jet" represents the quark or gluon in the final state.
The solid histogram shows the distribution of the fragmentation events simulated by PYTHIA.
The dashed histogram shows the distribution of the $qg \rightarrow \gamma \gamma + q$ events 
generated by GR@PPA in which the initial-state and final-state LLL subtractions are fully applied. 
In both simulations, the event selection has been applied to those events fully simulated down to the hadron level, 
while $\Delta R(\gamma\text{-jet})$ has been derived from the information of the reconstructed or original 
$\gamma \gamma + \text{jet}$ events at the on-shell parton level.
The sum of the two distributions is plotted with filled circles.
The dotted histogram shows the distribution directly derived from the $qg \rightarrow \gamma \gamma + q$ matrix element 
before applying the final-state QED LLL subtraction (same as the solid histogram in Fig.\ \ref{fig:drgamq}).
The distribution of the $q\bar{q} \rightarrow \gamma \gamma + g$ events is shown with a dot-dashed 
histogram for comparison.
\protect\label{fig:drgamj-noiso}}
\end{figure}

The $\Delta R(\gamma\text{-jet})$ distribution of the simulated fragmentation events is shown 
with a solid histogram in Fig.\ \ref{fig:drgamj-noiso}, 
in which the selection condition in Eq.\ (\ref{eq:selection}) was applied to the simulated hadron-level events, 
while $\Delta R(\gamma\text{-jet})$ was evaluated using the reconstructed on-shell parton level 
$qg \rightarrow \gamma \gamma + q$ information.
In order to compare with this result, 
we applied a full simulation down to the hadron level to the GR@PPA simulation events described in the previous section.
The events were generated with a looser kinematical constraint,
 $p_{T}(\gamma_{1}) \geq 30 \text{ GeV/}c, p_{T}(\gamma_{2}) \geq 20 \text{ GeV/}c, \text{ and } |\eta(\gamma)| \leq 3.0$, 
and the initial-state and final-state parton showers were applied down to $Q_{0}^{2} = (4.6 \text{ GeV})^{2}$ in GR@PPA.
The generated events were passed to PYTHIA in order to simulate parton showers at smaller $Q^{2}$ and hadronization/decays.
The default setting was unchanged in the PYTHIA simulation, except for the setting of {\tt parp(67) = 1.0} and {\tt parp(71) = 1.0}.
The selection condition in Eq.\ (\ref{eq:selection}) was then applied to the simulated hadron-level events.
The $\gamma$-jet separation, $\Delta R(\gamma\text{-jet})$, was evaluated from the original on-shell parton level information 
for the selected events.
The obtained $\Delta R(\gamma\text{-jet})$ distribution is presented with a dashed histogram in Fig.\ \ref{fig:drgamj-noiso}.

The obtained two distributions seem to be smoothly connected with some overlap around the boundary.
We expect that the sum of them should reproduce the spectrum before the final-state subtraction 
shown with a dotted histogram around the boundary, 
since we have chosen the boundary, $Q^{2} = \mu_{F}^{2}$, in a hard radiation region.
The sum should gradually get apart from the prediction without the subtraction 
to approach the prediction from the fragmentation as $\Delta R(\gamma\text{-jet})$ becomes smaller.
The sum shown with closed circles in Fig.\ \ref{fig:drgamj-noiso} behaves nearly as expected.
However, we can see a small dip at the boundary. 
This dip must be due to a certain mismatch in the hard radiation kinematics.
Furthermore, the distribution of the fragmentation events seems to be a little bit smaller than we expect.
The distribution looks better if we enhance the fragmentation by a factor of about 20\%.
We expect that we can achieve a better matching with the parton shower that we are developing.

People may worry that the drop in the first bin of the fragmentation event distribution 
in Fig.\ \ref{fig:drgamj-noiso} looks unnatural.
This drop is caused by a cutoff of the QED PS in PYTHIA.
The cutoff is set to 1 GeV in terms of $Q$ by the default, to be equal to that of the QCD PS.
Possible QCD phenomena at smaller $Q$ values are simulated otherwise, for instance, by the multiple interaction,
while no such simulations are implemented for QED.
Thus, there is no reason to stop the QED PS at $Q =$ 1 GeV.
Hard photons can be radiated from branches having smaller $Q$ values 
since the momentum determination is independent of the choice of $Q$.
Actually, if we decrease this cutoff to 0.3 GeV by setting as {\tt parj(83) = 0.3} in PYTHIA, 
the concentration in the first bin increases significantly.
Though the concentration will further increase if we apply smaller cutoff values, 
there must be a certain limitation due to non-perturbative effects of QCD. 
The picture in which a quark branches to a quark and a photon should break at $Q$ values smaller than the QCD cutoff.
The treatment of such small-$Q^{2}$ branches is one of the subjects in our development.

In Fig.\ \ref{fig:drgamj-noiso}, the contribution from $q\bar{q} \rightarrow \gamma \gamma + g$ events 
is shown with a dot-dashed histogram for comparison.
The initial-state QCD LLL subtraction is applied also to this process.
This process does not have any final-state divergence.
We can see that the contribution from this process is always smaller than  $qg \rightarrow \gamma \gamma + q$ 
by nearly a factor of five.

\section{Isolation cut}

\begin{figure}[t]
\centerline{\psfig{file=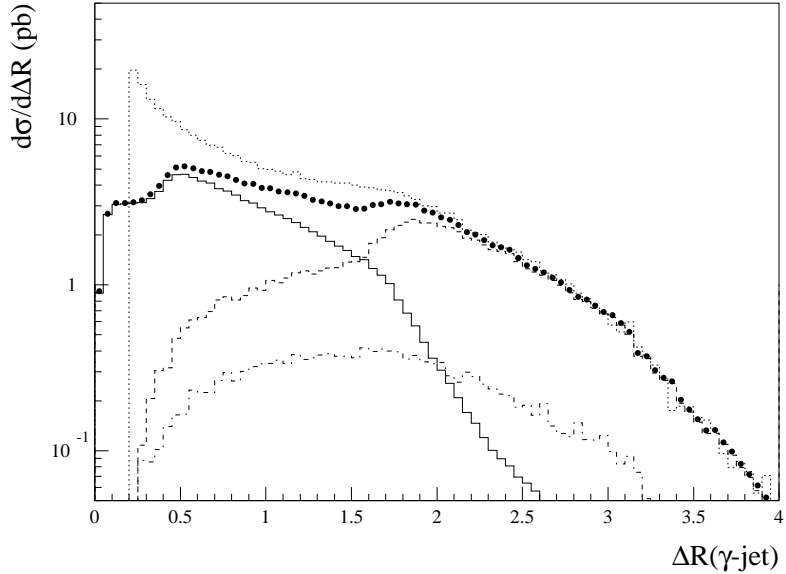,width=120mm}}
\caption{$\Delta R(\gamma\text{-jet})$ distribution of the simulated $\gamma \gamma + \text{jet}$ events.
The isolation cut described in the text is applied to the events 
from which the distributions in Fig.\ \ref{fig:drgamj-noiso} are derived.
The notation of the histograms and the plot is the same as Fig.\ \ref{fig:drgamj-noiso}.
\protect\label{fig:drgamj-iso}}
\end{figure}

It is necessary in real experiments to require a certain isolation condition in the identification of photons, 
in order to reduce the huge background from $\pi^{0}$ in hadron jets.
It is difficult to reproduce the cuts applied by experiments with parton-level simulations.
This is the main reason why hadron-level event generators are desired.
If hadron-level events can be generated consistently, 
the generated events can be passed to detector simulations for more detailed studies including detector responses.

\begin{figure}[t]
\centerline{\psfig{file=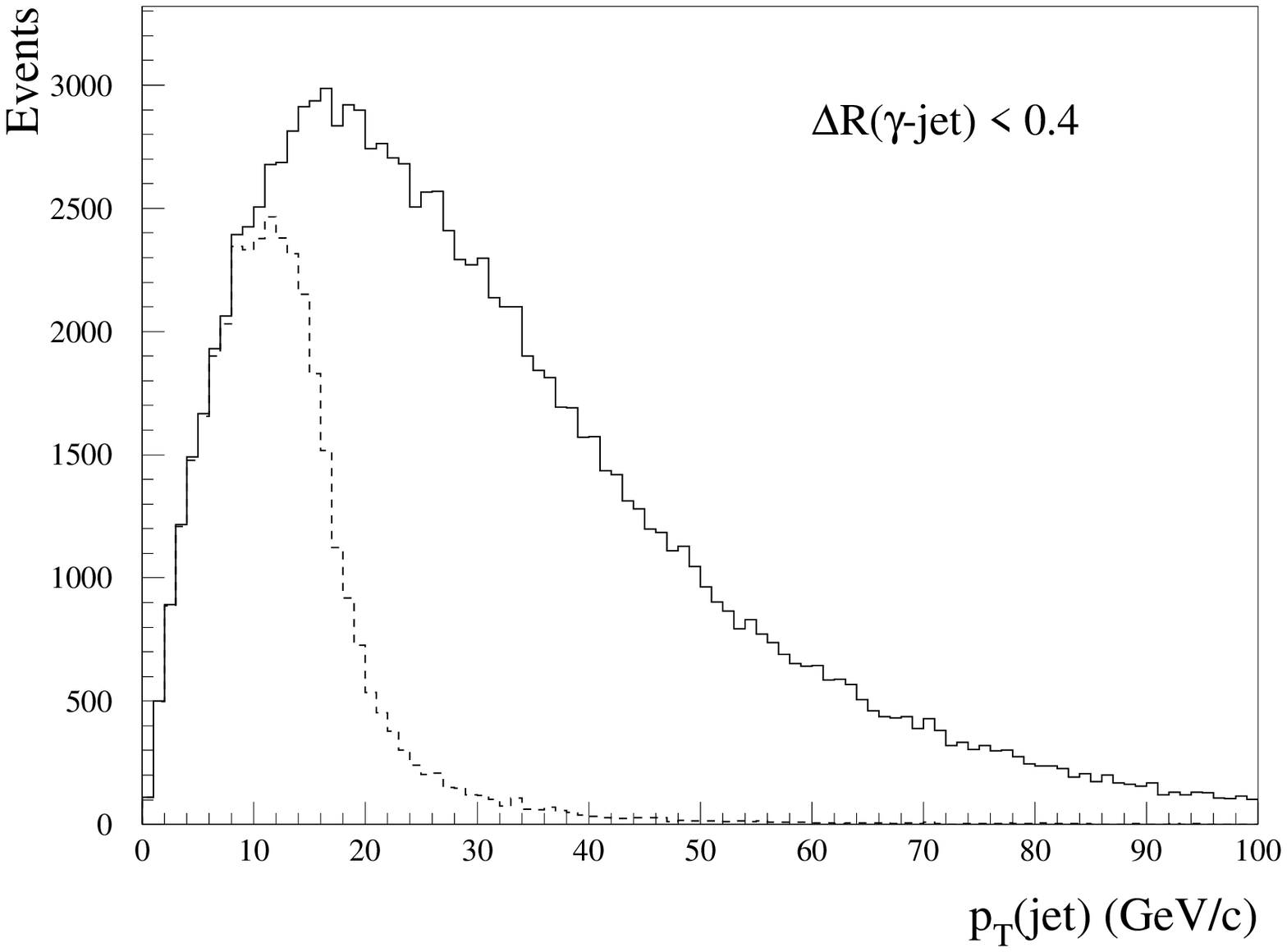,width=120mm}}
\caption{$p_{T}$ distribution the "jets" in the on-shell parton level $\gamma \gamma + \text{jet}$ events 
reconstructed for the fragmentation events in $\Delta R(\gamma\text{-jet}) \leq 0.4$.
The solid histogram shows the distribution before the isolation cut, and the dashed histogram shows the 
distribution after the cut, $E_{T,\text{cone}} \leq$ 15 GeV.
\protect\label{fig:ptjet}}
\end{figure}

In this section we try to simulate a typical isolation condition using the events simulated down to the hadron level.
The condition that we require is defined by using a cone $E_{T}$ defined as
\begin{equation}\label{eq:econe}
  E_{T,\text{cone}} = \sum_{\Delta R < R_{\text{iso}} \text{ wrt. } \gamma}E_{T},
\end{equation}
where $E_{T}$ is the transverse component of the energy of particles with respect to the beam direction, 
and the sum is taken over all particles inside a given $\Delta R$ cone around the photon, 
excluding the photon under the study and neutrinos.
We take as $R_{\text{iso}} = 0.4$ and require that $E_{T,\text{cone}} \leq$ 15 GeV.
We applied this cut to the events from which the distributions in Fig.\ \ref{fig:drgamj-noiso} were derived.
The $\Delta R(\gamma\text{-jet})$ distribution after applying the cut is shown in Fig.\ \ref{fig:drgamj-iso}.
We can see that the isolation cut strongly suppresses the fragmentation events 
in $\Delta R(\gamma\text{-jet}) \lesssim 0.4$, as expected, 
while it has a very small impact to the LLL subtracted $qg \rightarrow \gamma \gamma + q$ 
and $q\bar{q} \rightarrow \gamma \gamma + g$ events.

Figure\ \ref{fig:ptjet} shows the $p_{T}$ distribution of the "jets" in the on-shell parton level 
$\gamma \gamma + \text{jet}$ events reconstructed for the fragmentation events 
in the region, $\Delta R(\gamma\text{-jet}) \leq 0.4$.
The distribution after the cut should sharply drop at 15 GeV/$c$ 
if the isolation cut at the on-shell parton level could perfectly reproduce the cut applied to the hadron-level events.
We can see an apparent smearing around the ideal cut value. 
This smearing would result in an inaccuracy of the approximation in analytical evaluations, such as DIPHOX and ResBos.
In any case, the fact that the distribution drops around the ideal cut value, as we expect, 
must imply that the "reconstruction" of the on-shell parton level events is reasonably done.

\section{Combined simulation}

We can obtain a consistent simulation sample by combining the fragmentation events 
with the LLL subtracted $qg \rightarrow \gamma \gamma + q$ and $q\bar{q} \rightarrow \gamma \gamma + g$ events, 
together with the lowest-order $q\bar{q} \rightarrow \gamma \gamma$ events.
The lowest-order events were simultaneously generated by GR@PPA with the LLL subtracted events, 
to restore the subtracted initial-state divergent components.
A loose kinematic condition, 
$p_{T}(\gamma_{1}) \geq 30 \text{ GeV/}c, p_{T}(\gamma_{2}) \geq 20 \text{ GeV/}c, \text{ and } |\eta(\gamma)| \leq 3.0$, 
was commonly applied in the on-shell parton level event generation in GR@PPA.
The event generation was done for the LHC design condition, $pp$ collisions at 14 TeV, 
with the energy scale definition in Eq.\ (\ref{eq:scale}).
Built-in parton shower simulations were applied in GR@PPA down to $Q_{0}^{2} = (4.6 \text{ GeV})^{2}$.
Further small-$Q^{2}$ phenomena were simulated by PYTHIA 6.423, with its default setting
but {\tt parp(67) = 1.0} and {\tt parp(71) = 1.0}.
The selection condition in Eq.\ (\ref{eq:selection}) was applied to thus simulated hadron-level events.
The fragmentation events were separately simulated with PYTHIA as described in previous sections.
Finally, the isolation cut described in the previous section was applied to the simulated events.

\begin{figure}[t]
\centerline{\psfig{file=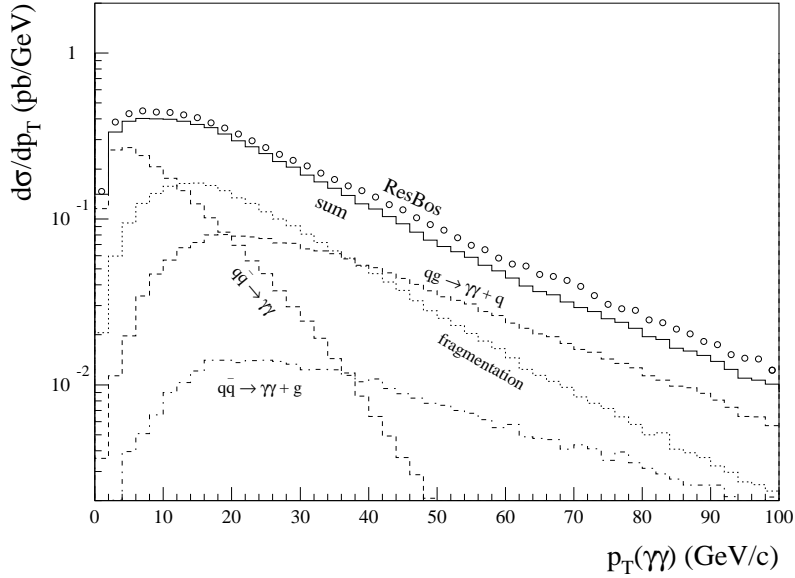,width=120mm}}
\caption{$p_{T}$ distribution of the diphoton ($\gamma\gamma$) system in the simulated events 
satisfying the kinematical condition in Eq.\ (\ref{eq:selection}) and the isolation condition.
The sum (solid histogram) is presented together with the distributions of the subprocesses: 
$q\bar{q} \rightarrow \gamma \gamma$, the LLL subtracted $qg \rightarrow \gamma \gamma + q$ 
and $q\bar{q} \rightarrow \gamma \gamma + q$ and the fragmentation.
The ResBos prediction for the same condition is plotted with open circles for comparison.
\protect\label{fig:ptgamgam}}
\end{figure}

Figure\ \ref{fig:ptgamgam} shows the $p_{T}$ distribution of the diphoton ($\gamma\gamma$) system 
in the simulated events.
The obtained distribution is compared with the prediction from ResBos \cite{Balazs:2007hr}.
ResBos provides us with an NLO prediction in which soft QCD radiations are resummed. 
It is considered to be most reliable at present as far as the $p_{T}(\gamma\gamma)$ distribution is concerned.
Unfortunately, since the resummation can predict inclusive properties only, 
ResBos cannot provide exclusive event information.
It should be noted that the contribution from the $gg \rightarrow \gamma\gamma$ process is not included 
in the ResBos prediction since it is yet to be included in our simulation.
The ResBos prediction shown in the figure is smaller than the result presented in the original paper 
because the $m_{\gamma\gamma}$ constraint in Eq.\ (\ref{eq:selection}) is additionally required.
We have confirmed that we can obtain a prediction close to the original result 
if the $m_{\gamma\gamma}$ constraint is excluded.

We can see that the sum (solid histogram) of our simulations is in reasonable agreement with the ResBos prediction, 
not only in the $p_{T}(\gamma\gamma)$ dependence but also in the absolute value of the cross section.
The total cross section is 15.5 pb from ResBos and 13.7 pb from our simulation.
The difference can be attributed mainly to the lack of finite terms in the NLO correction in our simulation.
The technical difference in the isolation cut may also have resulted in some difference.

In our simulation, the contribution from each subprocess is: 
4.1 pb (30\%) from $q\bar{q} \rightarrow \gamma \gamma$, 5.1 pb (37\%) from the fragmentation, 
and 3.7 pb (27\%) and 0.8 pb (11\%) from the LLL subtracted $qg \rightarrow \gamma \gamma + q$ 
and $q\bar{q} \rightarrow \gamma \gamma + g$ processes, respectively.
The contribution from the lowest-order process, $q\bar{q} \rightarrow \gamma \gamma$, 
is smaller than 1/3 of the total sum.
Furthermore, as we can see in Fig.\ \ref{fig:ptgamgam}, the $p_{T}(\gamma\gamma)$ spectrum of 
$q\bar{q} \rightarrow \gamma \gamma$ is apparently different from the sum.
The contribution from the LLL subtracted $qg \rightarrow \gamma \gamma + q$ 
is comparable with the lowest-order contribution.
However, it must be noted that the composition in the above is not physically meaningful.
The above is a result when we separate the soft and hard radiations at $\mu = p_{T}$ 
of the $q\bar{q} \rightarrow \gamma \gamma$ or $qg \rightarrow \gamma + q$ interaction.
The composition changes if we adopt another definition.
In any case, we can at least conclude that the contribution from the $qg \rightarrow \gamma \gamma + q$ process 
in which the final-state three particles are well separated is significant in the non-resonant QED diphoton production.
Together with the LL component which is shared between the fragmentation process and the LLL subtracted process, 
non-LL components remaining after the subtraction also looks sizable 
as we can see in Figs.\ \ref{fig:drgamj-noiso} and \ref{fig:drgamj-iso}. 

\section{Conclusion}

In conclusion, we have developed an exclusive event generator for non-resonant QED diphoton ($\gamma\gamma$) 
production at hadron collisions, consistently including additional jet productions. 
The $qg \rightarrow \gamma \gamma + q$ process to be included as a radiative process 
has a final-state QED divergence as well as an initial-state QCD divergence.
We have developed a subtraction method to regularize the final-state divergence, 
by extending the method developed for initial-state QCD divergences. 
The subtraction works well as we expected.
The differential cross section converges to zero after the subtraction 
at the limit where one of the pairs of the final-state $\gamma$ and quark becomes collinear 
and the original cross section diverges.

We tried to use the PYTHIA simulation for the generation of the fragmentation events, 
with which the subtracted component is to be restored without any double count.
We generated $\gamma$ + jet events by PYTHIA and picked up another photon 
produced by the parton shower (PS) in the final state.
We have developed a technique to reconstruct on-shell parton level $\gamma\gamma$ + jet events 
from those events fully simulated down to the hadron level, 
in order to test the matching with the events generated by GR@PPA.
The fragmentation events generated by using the default "old" PS show a reasonable matching, 
though a small mismatch is seen around the boundary and the overall yield seems to be a little bit smaller.
The treatment of small-$Q^{2}$ radiations is still an open question in this simulation.

A consistent simulation sample was composed by combining the above two generation samples 
with the $q\bar{q} \rightarrow \gamma \gamma$ and $q\bar{q} \rightarrow \gamma \gamma + g$ events 
simultaneously generated with $qg \rightarrow \gamma \gamma + q$ in GR@PPA.
The event generation was tested for the LHC design condition, $pp$ collisions at 14 TeV, 
and a typical Higgs-boson search condition was required to the events simulated down to the hadron level.
A typical isolation cut was also applied to the photons.
We observed a reasonable suppression of the fragmentation events in a collinear region, 
together with a visible smearing of the boundaries due to the application of PS and hadronization.

The combined events show a behavior which reasonably agrees with the prediction from 
a resummed NLO calculation by ResBos.
The difference in the total cross section at the level of 10\% can be attributed to the lack of 
finite-term contributions in the NLO correction in our simulation.
In our simulation sample, for which all the energy scales were chosen to be equal to the $p_{T}$ of 
the basic $q\bar{q} \rightarrow \gamma \gamma$ or $qg \rightarrow \gamma + q$ process, 
the contribution from the lowest-order $q\bar{q} \rightarrow \gamma \gamma$ process is 
smaller than 1/3.
The contribution from $qg \rightarrow \gamma \gamma + q$ is comparable with it 
due to a large density of gluons inside protons.
A consistent inclusion of this process must be necessary for reliable studies. 

A new PS which implements QED radiations together with QCD is under development.
We want to implement a mechanism to force a hard photon radiation in this PS 
in order to ensure a reasonable generation efficiency for the fragmentation process. 

\section*{Acknowledgments}

This work has been carried out as an activity of the NLO Working Group, 
a collaboration between the Japanese ATLAS group and the numerical analysis 
group (Minami-Tateya group) at KEK.
The author wishes to acknowledge useful discussions with the members, 
especially continuous discussions with Y. Kurihara.

%\section*{References}

\end{document}